\documentclass[twocolumn,floats,prl,superscriptaddress,showpacs,aps]{revtex4}

\usepackage[dvips]{graphicx}

\begin{document}

\title{Infrared Properties of Electron Doped Cuprates: Tracking Normal State 
       Gaps and Quantum Critical Behavior in Pr$_{2-x}$Ce$_{x}$CuO$_4$ }
       
\author{A. Zimmers}
\author{J.M. Tomczak}
\author{R.P.S.M. Lobo}
\email{lobo@espci.fr}
\author{N. Bontemps}
\affiliation{Laboratoire de Physique du Solide (UPR 5 CNRS) ESPCI, 10 rue 
Vauquelin 75231 Paris, France.}

\author{C.P. Hill}
\author{M.C. Barr}
\author{Y. Dagan}
\author{R.L. Greene}
\affiliation{Center for Superconductivity Research, Department of Physics, 
University of Maryland, College Park, Maryland 20742, USA.}

\author{A.J. Millis}
\affiliation{Physics Department, Columbia University, New York, New York 10027, 
USA.}

\author{C.C. Homes}
\affiliation{Department of Physics, Brookhaven National Laboratory, Upton, New 
York 11973, USA.}

\date{\today}

\begin{abstract}
We report the temperature dependence of the infrared-visible conductivity of 
Pr$_{2-x}$Ce$_x$CuO$_{4}$ thin films. When varying the doping from a 
non-superconducting film ($x=0.11$) to a superconducting overdoped film 
($x=0.17$), we observe, up to optimal doping ($x=0.15$), a {\it partial} gap 
opening. A model combining a spin density wave gap and a frequency and 
temperature dependent self energy reproduces our data reasonably well. The 
magnitude of this gap extrapolates to zero for $x \sim 0.17$ indicating the 
coexistence of magnetism and superconductivity in this material and the 
existence of a quantum critical point at this Ce concentration.
\end{abstract}


\maketitle 

Over the last 15 years significant work has been done on the differences and 
similarities between electron and hole-doped cuprates \cite{FournierReview}. The 
two material families share a structure with CuO$_2$ planes, both exhibit 
superconductivity in a moderate doping range with $\sim 0.16$ carriers per 
Copper, and both exhibit anomalous `normal' (non-superconducting) states 
characterized in some doping and temperature ranges by a normal state gap or 
`pseudogap'.

In hole-doped materials early evidence for a `pseudogap' came from nuclear 
magnetic resonance experiments \cite{Alloul89,Takigawa91}. Angle-resolved 
photoemission (ARPES) experiments performed on the Bi-2212 material show that 
the pseudogap opens along the $(0,\pi)$ direction in $k$ space \cite{Norman} 
and evolves smoothly into the superconducting gap as the temperature $T$ is 
lowered. However, in the hole-doped materials gap-like features do not appear
in the the optical conductivity and the pseudogap is only related to a sharp 
decrease in the optically defined scattering rate \cite{Puchkov} instead of an 
upward shift in spectral weight for a density wave gap \cite{Santander-Syro}.

On the electron-doped side, the most studied material is 
Nd$_{2-x}$Ce$_x$CuO$_{4}$ (NCCO) \cite{Homes1997, Singley2001}. The optical 
conductivity of non superconducting single crystals ($x=0$ to $0.125$) shows the 
opening of a high energy ``pseudogap'' at temperatures well above the $T_{Neel}$ 
associated with antiferromagnetic ordering \cite{Onose}. ARPES measurements 
mapping the Fermi surface at low temperature for $x=0.04$, $0.10$ and $0.15$ 
suggest the presence of pockets, as expected from long ranged magnetic order. 
For $x=0.15$ intensity is suppressed where the nominal Fermi surface crosses the
magnetic Brillouin zone boudary \cite{Armitage}. 

In this Letter we report measurements of the temperature evolution of the 
optical conductivity in a set of  Pr$_{2-x}$Ce$_{x}$CuO$_{4}$ (PCCO) thin films. 
The PCCO material has a wider superconducting range than does NCCO. Thin films 
are extremely homogeneous in the Ce concentration and are easier to anneal than 
crystals. Most important they can be made superconducting in the underdoped 
regime, whereas this seems difficult for crystals \cite{Onose}. Our new 
generation of films are large enough to allow accurate optical studies, enabling 
us to track the optical behavior to lower energy scales and into the 
superconducting state \cite{Zimmers}. Our data reveals the onset of a ``high 
energy'' partial gap below a characteristic temperature $T_W$ which evolves with 
doping. The gap is directly evident in the measured optical conductivity for 
$x=0.11$ and $0.13$, it is absent down to $5$ K for $x=0.17$ and it has a subtle 
signature for $x=0.15$ (optimal doping). This partial gap is present in our 
superconducting samples in contrast to ref. \cite{Onose}. The closure of the gap 
suggests a quantum critical point (QCP) around $x = 0.17$, consistent with 
transport evidence on similar samples \cite{Dagan}. We suggest that this gap 
originates from a spin density wave (SDW), consistent with ARPES 
\cite{Armitage}, and that the QCP is an antiferromagnetic-paramagnetic one.

The thin films studied in this work were epitaxially grown by pulsed-laser 
deposition on a SrTiO$_3$ substrate \cite{FournierSample}. The samples studied 
are (i) $x=0.11$, not superconducting down to $4$ K (thickness 2890 \AA), (ii) 
$x=0.13$ (underdoped) $T_c=15$ K (thickness 3070~\AA), (iii) $x=0.15$ (optimally 
doped), $T_c=21$ K (thickness 3780~\AA), (iv) $x=0.17$ (overdoped) $T_c=15$ K 
(thickness 3750~\AA). $T_c$ for all the films were obtained by electrical 
resistance measurements. Infrared-visible reflectivity spectra were measured for 
all samples in the 25--21000~cm$^{-1}$ spectral range with a Bruker IFS 66v 
Fourier Transform spectrometer within an accuracy of 0.2\%. Data was taken at 
typically 12 temperatures (controlled better than 0.2 K) between 25 K and 300 K. 
The far-infrared (10--100~cm$^{-1}$) was measured for samples (iii) and (iv) 
utilizing a Bruker IFS 113v at Brookhaven National Laboratory. 

\begin{figure}
  \begin{center}
	\includegraphics[width=7cm]{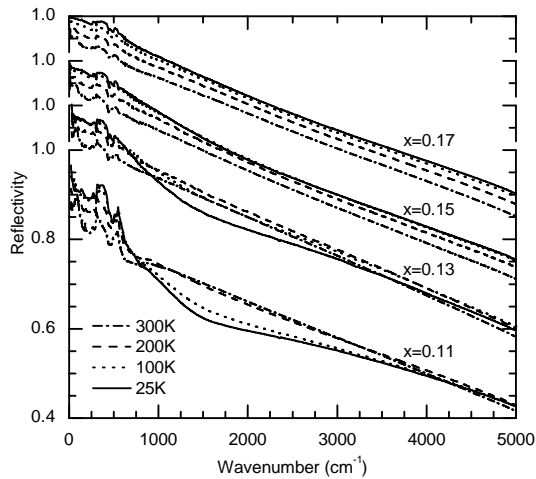}
  \end{center}
\caption{Raw infrared reflectivity of $x=0.11$, 0.13, 0.15 and 0.17 samples. 
Curves for different concentrations are shifted from one another by 0.1 for 
clarity.} 
\label{fig1}
\end{figure}

Figure \ref{fig1} shows the raw reflectivity from 25 to 5000 cm$^{-1}$ for a set 
of selected temperatures. As the temperature decreases, a suppression of $R$ 
becomes conspicuous for $x=0.11$ and 0.13 and is visible for $x=0.15$ as a 
bending of the lowest temperature curve, crossing at 75 K. Conversely, the 
reflectivity of the $x=0.17$ sample increases monotonically with decreasing 
temperature over the whole spectral range shown. 

\begin{figure}
  \begin{center}
    \includegraphics[width=8cm]{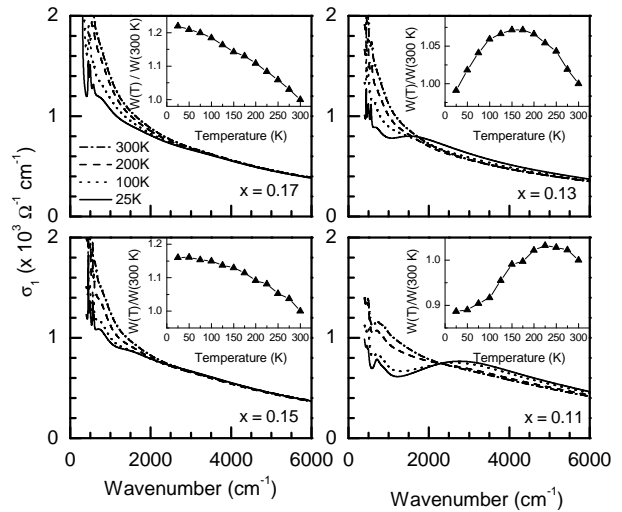}
  \end{center}
\caption{Real part of the optical conductivity from 400 to 6000 cm$^{-1}$. The 
inset in each panel shows the normalized spectral weight 
$W(\omega_H,T) / W(\omega_H, 300 \textrm{ K})$ ($\omega_L = 0$) plotted 
{\it vs.} temperature for an upper cut-off frequencies 
$\omega_H = 2000 \textrm{ cm}^{-1}$.}
\label{fig2}
\end{figure}

We applied a standard thin film fitting procedure to extract the optical 
conductivity from this data set \cite{Santander-Syro, Santander-these}. The real 
part $\sigma_1(\omega)$ of the optical conductivity is plotted in Fig. 
\ref{fig2}. At low energy, for all concentrations, there is a Drude-like 
contribution which narrows as the temperature is lowered in the normal state 
from 300~K to 25~K. This implies a transfer of spectral weight, from higher 
(above, e.g., $\approx 170$ cm$^{-1}$ for $x=0.13$) to lower frequencies. Above 
1000~cm$^{-1}$, the feature noticed in the raw reflectivity for $x=0.11$ and 
0.13 produces a dip/hump structure. $\sigma_1 (\omega)$ peaks at 
$\Omega_M \sim 2750$~cm$^{-1}$ for $x=0.11$ and at $\sim 1500$~cm$^{-1}$ for 
$x=0.13$. A similar feature was observed in NCCO single crystals only for doping 
levels where such crystals are {\it not superconducting} \cite{Onose}, in 
contrast to our observation in the $x=0.13$ sample. 

\begin{figure}
  \begin{center}
    \includegraphics[width=8cm]{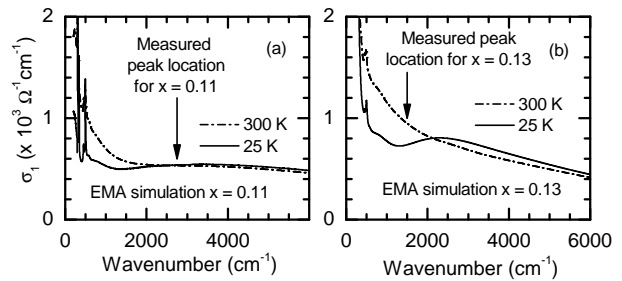}
  \end{center}
\caption{Effective medium approximation (EMA) simulations of conductivity. {\it 
Left panel}: attempts to simulate $x=0.11$ crystal by combinations of undoped 
and $x=0.17$ conductivities. {\it Right panel}: attempts to simulate 
conductivity of $x=0.13$ sample by combinations of $x=0.11$ and $x=0.17$. The 
proportions of each phase were chosen so that the average Ce concentration is 
the nominal one.}
\label{fig3}
\end{figure}

Homogeneity issues are a key question in cuprates \cite{Davis}. We investigated 
the homogeneity of our samples in several ways. Using the X-ray analysis of an 
EDAX system, we verified that the $x=0.15$ sample did not present any trace of 
inhomogeneity on the micron scale in the Pr, Ce or Cu concentrations. We also 
used the standard Bruggeman effective medium approximation \cite{EMA} (valid for 
inhomogeneities smaller than the wavelength but larger than the mean free path 
of $\lesssim 50$ \AA) to investigate the possibility that the gap features 
observed in one concentration could arise from an inhomogenous mixture of two 
different concentrations. The left hand panel of Fig. \ref{fig3} shows results 
of an attempt to simulate the optical conductivity of the $x=0.11$ sample as a 
combination of 35\% x=0 \cite{Homes} and 65\% x=0.17, as suggested in Ref. 
\cite{Uefuji}. The failure to describe the spectral weight shift and peak 
development is evident. The right panel of Fig. \ref{fig3} shows results of an 
attempt to simulate the $x=0.13$ sample from the $x=0.11$ and $x=0.17$ ones. For 
no parameters were we able to place the peak in $\sigma_1$ at the correct 
location. These, and other simulations (not shown) lead us to believe that the 
structures we observe are intrinsic.

We now present a spectral weight analysis which shows that the features seen in 
$\sigma_1$ and $R$ are due to the opening of a partial gap. We define the 
restricted spectral weight $RSW(\omega_L,\omega_H,T)$ via
\begin{equation}
RSW(\omega_L,\omega_H,T) = 
  \frac{2}{\pi} \int_{\omega_L}^{\omega_H} \sigma_1(\omega,T)d\omega.
\label{eq1}
\end{equation}
The familiar $f$-sum rule implies $RSW(0,\infty,T) = n e^2 / m$; the partial 
integrals provide insight into the rearrangement of the conductivity with 
temperature. In a conventional metal the conductivity is decribed by a ``Drude 
peak'' centered at $\omega = 0$ and with half width $\Gamma$ decreasing as $T$ 
is lowered. In this case spectral weight is transfered from high to low energies 
as $T$ decreases. On the other hand, the opening of a density wave gap leads to 
a transfer of spectral weight to higher energies, beyond the gap edge.

In all of our samples, $RSW(0,20000\textrm{ cm}^{-1},T)$ is temperature 
independent within 3\%; however the partial integrals display an informative 
$T$-dependence. The insets to Fig. \ref{fig2} display 
$W(\omega_H,T) = RSW(0,\omega_H,T)$ for $\omega_H = 2000 \textrm{ cm}^{-1}$, all 
normalized to the $T=300$ K values. The presence of a narrow Drude peak 
($\lesssim 100 \textrm{cm}^{-1}$) implies that accurate low frequency data 
(measured down to $10 \textrm{cm}^{-1}$) is important in order to get reliable 
zero-frequency extrapolations. The $x = 0.17$ curves display the steady increase 
in $W$ expected of a Drude metal. On the other hand, both the $x=0.11$ and 
$x=0.13$ samples display a non-monotonic behavior indicating an upward shift of 
spectral weight beginning below $T \approx 225$ K ($x = 0.11$) and 
$T \approx 150$ K ($x=0.13$); however we note that all samples remain metallic 
at low temperatures as indicated by the presence of a Drude peak. In drawing 
this conclusion it is important to integrate down to zero to insure that the 
weight is not transferred downwards into a narrow Drude peak. We believe that 
the only consistent interpretation of these data is that as $T$ is lowered a gap 
appears on {\it part} of the Fermi surface. In particular, simulations of two 
component (Drude and mid-infrared) models, such as the polaron model of Lupi 
{\it et al.} \cite{Lupi} leads to negligible upwards transfer of spectral weight 
($\sim 5$ \% of observed values).

\begin{figure}
  \begin{center}
    \includegraphics[width=8cm]{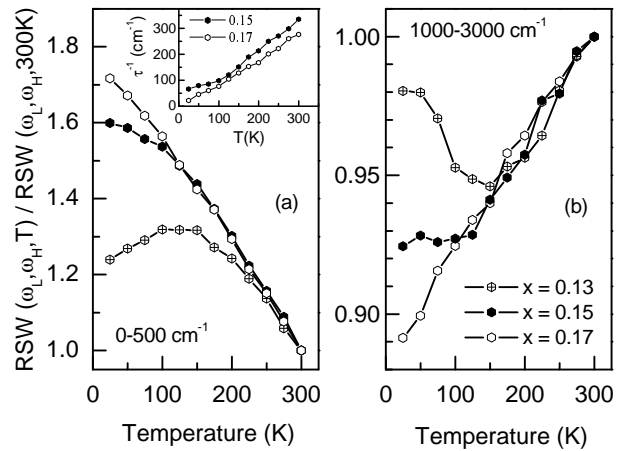} 
  \end{center}
\caption{Temperature evolution of the restricted spectral weight. The 
integration boundaries $\omega_L$--$\omega_H$ are indicated in each panel. The 
inset shows the dc extrapolation for the scattering rate for the $x=0.15$ and 
0.17 samples.}
\label{fig4}
\end{figure}

To address the issue of the existence of a partial gap in the $x=0.15$
(optimally doped) sample we compare in Fig. \ref{fig4} the $RSW$ of $x=0.13$, 
0.15 and 0.17 in two frequency ranges [{\it panel (a)}: 0 to 500 cm$^{-1}$ and 
{\it panel (b)}: 1000 to 3000 cm$^{-1}$). For clarity, each curve is normalized 
to its value at 300 K. In both panels, the $x=0.17$ sample displays a monotonic 
evolution, as expected from the temperature dependence of its scattering rate 
shown in the inset. The $x=0.13$ sample displays a non-monotonic behavior in 
both frequency ranges. In panel (a) the increase of spectral weight due to the 
Drude narrowing is overcome at low temperatures by the gap opening. As expected, 
this trend is reversed in panel (b). For the $x=0.15$ sample, the low $\omega$ 
$RSW$ [panel (a)] shows at low $T$ ($\lesssim 100$ K) a pronounced flattening 
relative to the $x = 0.17$ sample. The high $\omega$ $RSW$ [panel (b)] shows the 
complementary effect: the decrease with decreasing $T$ is halted below 
$T \approx 100$ K. On the other hand, the free carrier scattering rate (inset) 
shows a smooth decrease over the whole temperature regime. This, combined with 
the clearly interpretable behavior of the $x=0.13$ and 0.17 compounds, strongly 
suggests that a small gap opens below 100 K for $x=0.15$.

\begin{figure}
  \begin{center}
    \includegraphics[width=8cm]{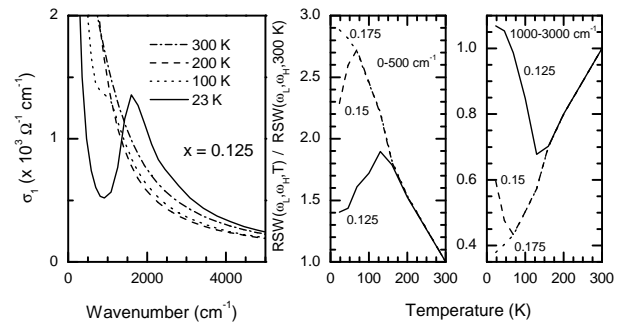} 
  \end{center}
\caption{{\it Left panel:} Optical conductivity calculated by a spin density 
wave model for $x=0.125$. {\it Mid and right panels:} RSW calculated for 
different dopings from the spin density wave model as described in the text.}
\label{fig5}
\end{figure}

A natural interpretation is that the observed optical gap arises from 
commensurate $(\pi ,\pi)$ magnetic order. Neutron scattering consistent with 
this order has been unambiguously observed at lower dopings 
\cite{Thurston1990,Matsuda92,Yamada99}, but whether the order exists at 
superconducting concentrations is not yet settled. To investigate whether this 
physics can lead to the gap observed in our experiments we have calculated the 
optical conductivity of a theoretical model of electrons moving in a band 
structure defined by the tight-binding dispersion appropriate to the cuprates 
\cite{Andersen95,Millis03} along with a $(\pi ,\pi)$ density wave gap of
magnitude $2 \Delta_{SDW}$ (corresponding to the photoemission band splitting) 
and mean field $T-$dependence. Our model uses the optical matrix elements 
appropriate to the tight binding model and also includes a frequency and 
temperature dependent self energy with imaginary part increasing from a 
$T$-dependent dc limit [chosen to roughly reproduce $\rho(T)$ at $T <T_W$] to a 
weakly $T$-dependent high $\omega$ limit. The specific form chosen is 
$\Sigma (\omega,T) = i \gamma_0 + 
  i \gamma_1 \left[ 1-\lambda(T)\omega_c\left( \omega_c - i\omega \right) /
  (\omega_c^2+\omega^2)\right] $
with $\gamma_0 = 0.01$ eV, $\gamma_1 = 0.25$ eV, $\omega_c = 0.3$ eV and 
$\lambda$ ranging from 0.7 at 300 K to 0.96 at low $T$. (Note that a 
`marginal Fermi liquid' self energy would have described the data almost as 
well, and leads to results very similar to those presented here.) As discussed 
in \cite{Millis03} we also included frequency and matrix element rescalings, 
by factors $\sim 0.6$ to account for high energy (Mott) physics. Thus the 
absolute values of $\sigma$ should be regarded as estimates, but the relative 
frequency and temperature dependence as well as spectral weight trends are
expected to be reliable. The calculated conductivity, for the doping $x=0.125$ 
and a gap which opens at 170 K and saturates at a $T=0$ value 
$2 \Delta_{SDW} = 0.14$ eV are shown in shown in the left panel of Fig. 
\ref{fig5}. The resemblance to the measured conductivity is striking. The right 
hand panels for Fig. \ref{fig5} show representative results of the modified 
spectral weight analysis for three dopings [$x=0.125$, $x=0.15$ ($T_W=70$ K, 
$2 \Delta_{SDW} = 0.03$ eV) and $x=0.175$ ($2\Delta_{SDW}=T_W=0$)]. The 
corresponding maximum in the optical conductivity occurs at 
$\Omega_M \approx 2.8 \Delta_{SDW}$. It is evident that the calculation 
reproduces the different qualitative behaviors of the restricted spectral 
weight. In particular, the minimum in the high frequency $RSW(T)$ curve 
corresponds to the opening of a gap. These curves therefore support the notion 
that the $x=0.15$ sample in fact has a small density wave gap, although this is 
not directly visible in the measured optical spectrum, and further supports that 
the temperature at which the gap opens may be inferred from the position of the 
minimum or saturation in the RSW.

\begin{figure}
  \begin{center}
    \includegraphics[width=6cm]{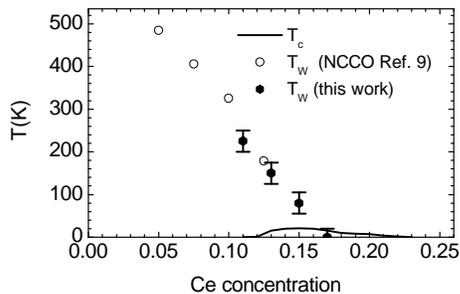} 
  \end{center}
\caption{$T_c$ and $T_{W}$, the SDW (or partial) gap opening temperature (see 
text) versus $x$. Open symbols are for NCCO from ref. \cite{Onose}, deduced from 
the maximum in $\sigma_1(\omega)$, and using the relation  
$\hbar \Omega_M / k_{B} T_W = 15$.}
\label{fig6}
\end{figure}

Detection of the gap by other means, especially via a low energy spectroscopy
which can probe the behavior as $\Delta \rightarrow 0$ in the $x=0.15-0.17$ 
range, would be very desirable. Intriguing tunneling measurements 
\cite{Biswas,Alff} suggest the presence of a gap which is, however, much smaller 
and closes at much lower temperature than the one we find.

Figure \ref{fig6} shows the gap onset temperature $T_W$ obtained from the 
breakpoints in Figs. \ref{fig2} and \ref{fig4}, along with those extracted from 
Onose {\it et al.} \cite{Onose} as a function of Ce concentration. Our work, 
covering a larger doping range up to the overdoped regime, strongly suggests 
that the $T_W$ line ends at a critical concentration of $x \sim 0.17$. This 
supports the arguments, obtained from dc transport in similar samples 
\cite{Dagan}, and strongly suggests that the QCP is of magnetic origin.  

In conclusion, we have measured with great accuracy the reflectivity of electron 
doped Pr$_{2-x}$Ce$_x$CuO$_{4}$ at various Ce doping levels ($x$). A careful 
optical conductivity spectral weight analysis shows that a partial gap opens 
below a temperature $T_W$ up to Ce concentrations of $x=0.15$. A spin density 
wave model reproduces satisfactorily the data where the gap has a clear spectral 
signature (for example $x=0.13$). The gap magnitude $2 \Delta_{SDW}$  relates to 
$T_W$  through $2 \Delta_{SDW} / k_B T_W \sim 11$. $T_W$ extrapolates to zero 
for $x \sim 0.17$, suggesting the presence of a quantum critical point inside 
the superconducting dome. We have shown that, in at least one class of 
high-$T_c$ superconductors, a `pseudogap' is associated with an an ordered phase 
terminating in a QCP at approximately optimal doping. We hope these results will 
provide a point of reference which will help to resolve similar issues arising 
in the hole-doped materials.

The authors thank Dr. V.N. Kulkarni for RBS / Channeling measurements and Dr. P. 
Bassoul for electronic microcopy studies. The work at University of Maryland was 
supported by NSF grant DMR-0102350. The work at Columbia was supported by NSF 
contract DMR-0338376. The Work at Brookhaven National Laboratory as supported by 
DOE under contract DE-AC02-98CH10886.


\begin{thebibliography}{00}

\bibitem{FournierReview}P. Fournier {\it et al.}, {\it in} The Gap Symmetry and 
Fluctuations in High-$T_c$ Superconductors, edited by J. Bok (Serie B: Physics 
Vol.371, NATO ASI Series, 1998).
\bibitem{Alloul89} H. Alloul, T. Ohno and P. Mendels, Phys. Rev. Lett. {\bf 63}
1700 (1989).
\bibitem{Takigawa91} M. Takigawa {\it et al.}, Phys. Rev. {\bf B43} 247 (1991).
\bibitem{Norman} M.N. Norman {\it et al.}, Nature {\bf 392}, 157 (1998).
\bibitem{Puchkov} A.V. Puchkov, D.N. Basov, and T. Timusk, J. Phys. Condens. 
Matter {\bf 8}, 10049 (1996).
\bibitem{Santander-Syro} A.F. Santander-Syro {\it et al.}, Phys. Rev. Lett. 
{\bf 88}, 097005 (2002).
\bibitem{Homes1997} C.C. Homes {\it et al.}, Phys. Rev. B {\bf 56}, 5525 (1997).
\bibitem{Singley2001} E.J. Singley {\it et al.}, Phys. Rev. B {\bf 64}, 224503 
(2001).
\bibitem{Onose} Y. Onose {\it et al.}, Phys. Rev. B {\bf 69}, 024504 (2004).
\bibitem{Armitage} N.P. Armitage {\it et al.}, Phys. Rev. Lett. {\bf 88}, 257001 
(2002).
\bibitem{Zimmers} A. Zimmers {\it et al.}, condmat/0405284.
\bibitem{Dagan} Y. Dagan {\it et al.}, Phys. Rev. Lett. {\bf 92}, 167001 (2004).
\bibitem{FournierSample} E. Maiser {\it et al.}, Physica C {\bf 297}, 15 (1998).
\bibitem{Santander-these} A.F. Santander-Syro {\it et al.}, condmat/0405264.
\bibitem{Davis} K. McElroy {\it et al.} Nature {\bf 413}, 282 (2001).
\bibitem{EMA} P. Wissmann and R.E. Hummel, {\it in} Handbook of optical 
properties. Vol. 2, edited by CRC Press (1997).
\bibitem{Homes} C. C. Homes {\it et al.}, Phys. Rev. B {\bf 66}, 144511 (2002).
\bibitem{Uefuji} T. Uefuji {\it et al.}, Physica C {\bf 392-396}, 189 (2003).
\bibitem{Lupi} S. Lupi {\it et al.}, Phys. Rev. Lett. {\bf 83}, 4852 (1999).
\bibitem{Thurston1990} T.R. Thurston {\it et al.}, Phys. Rev. Lett. {\bf 65}, 
263 (1990).
\bibitem{Matsuda92} M. Matsuda {\it et al.}, Phys. Rev. B {\bf 45}, 12548 (1992).
\bibitem{Yamada99} K. Yamada {\it et al.}, J. Phys. Chem. Sol. {\bf 60}, 1025 
(1999).
\bibitem{Andersen95} O.K. Andersen, cond-mat/9509044
\bibitem{Millis03} A.J. Millis and H.D. Drew, Phys. Rev. B {\bf 67}, 214517 
(2003).
\bibitem{Biswas} A. Biswas {\it et al.}, Phys. Rev. B {\bf 64}, 104519 (2001).
\bibitem{Alff} L. Alff {\it et al.} Nature {\bf 422}, 698 (2003).

\end{thebibliography}
\end{document}